\documentclass[aps, prb, twocolumn, longbibliography]{revtex4-2}
\usepackage{amsmath,amssymb}
\usepackage[dvipdfmx]{graphicx}
\usepackage{bm}
\usepackage{braket}
\usepackage{color}

\newcounter{num}

\begin{document}
\title{Antiparallel spin polarizations as quadratic response in chiral systems}

\author{Akane Inda, Kohei Hattori$^1$, Hiroaki Kusunose$^2$ and Satoru Hayami}
\affiliation{
Graduate School of Science, Hokkaido University, Sapporo 060-0810, Japan\\
$^1$Department of Applied Physics, The University of Tokyo, Bunkyo, Tokyo 113-8656, Japan\\
$^2$Department of Physics, Meiji University, Kanagawa 214-8571, Japan\\
}
\begin{abstract}
  Chirality-dependent spin generation has attracted considerable attention in condensed matter physics. 
  In this paper, we theoretically investigate antiparallel spin polarization as a chirality-dependent quadratic response, by using a finite chiral system composed of triangular prisms. 
  Based on the nonlinear Kubo formalism and real-time simulations, we demonstrate that spatially inhomogeneous antiparallel spin polarizations are induced as a dissipative quadratic DC response to a homogeneous AC electric field.
  In particular, we elucidate role of microscopic parameters characterizing the handedness of chirality, and naive expectation of spin polarization as a consequence of spin accumulation of spin current.
\end{abstract}

\maketitle
\section{Introduction}
Chirality is one of the fundamental aspects of nature~\cite{Bloom2024, Bousquet2025, Togawa2023, cheong2021permutable, cheong2022linking, kusunose2024emergence, Juraschek2025, Zhang2026}, arising in systems that lack both spatial inversion and mirror symmetries. 
This symmetry property is characterized by a time-reversal-even pseudoscalar~\cite{L.D.Barron_1986_true-chirality, Barron_mol-light-scattering}. 
Recent studies have increasingly focused on a microscopic description of chirality based on electronic degrees of freedom~\cite{Oiwa_PhysRevLett.129.116401, oiwa2025prr_te_se, inda2024quantification, hoshino2023spin-derived, Miki_PhysRevLett.134.226401}. 
Among such descriptions, the electric toroidal monopole (ETM) has been proposed as an electronic order parameter of chirality, on the basis of a complete multipole description~\cite{hayami2018prb_Classification_of_atomic-scale_multipoles, kusunose2020jpsj_completebasis, Kusunose_PhysRevB.107.195118, hayami2024unified,J.Kishine_IJC_2022_G0}. 
Such a multipole description enables us to extract the relevant electronic degrees of freedom --- in terms of charge, spin, orbital, and bond --- directly from a microscopic Hamiltonian. 
ETM degree of freedom is represented in various ways, for example, a spin-dependent imaginary hopping~\cite{Hayami_PhysRevLett.122.147602}, hybridization~\cite{hayami2023chiral, hayami2024analysis}, orbital-exchange hopping~\cite{oiwa2025prr_te_se}, or orbital cluster ordering~\cite{ishitobi2025purely}.  
This multipole-based framework further allows a quantitative evaluation of chirality directly from electronic wave functions, and has been successfully applied to a variety of chiral materials, such as Te~\cite{Oiwa_PhysRevLett.129.116401, oiwa2025prr_te_se}, Se~\cite{oiwa2025prr_te_se}, and a twisted methane molecule~\cite{inda2024quantification}.

Chirality has also attracted significant interest, as it gives rise to a wide range of intriguing physical properties and phenomena via a characteristic coupling between axiality and polarity degrees of freedom~\cite{cheong2021permutable, cheong2022linking, kusunose2024emergence, hayami2024unified, hayami2025chirality}.
A prominent example is the chirality-induced spin selectivity (CISS) effect~\cite{B.Gohler_nat_2011_CISS, Xie_2011_CISS, O.Ben_nat_2017_CISS, K.Michaeli_PNAS_2019_CISS, Suda_natcom_2019_CISS, Haipeng_2019_CISS, Naaman2020_2020_chiral_molecules_spin_selectivity, A.Inui2020prl_CrNb3S6, R.Neeman_ACR_2020_CISS, D.H.Waldeck2021aplmat_CISS, K.Shiota2021prl_disilicide_CISS,F.Evers_admat_2022_CISS, A.Kato2022prb_CISS, Nakajima23, K.Ohe_2024_quartz_CISS, Bloom_chemrev_2024_CISS}, in which electron spins are highly polarized when electrons traverse chiral materials. 
Although the microscopic origin of CISS remains under active debate, such spin-polarized states have been observed in a wide range of chiral systems, ranging from organic molecules~\cite{B.Gohler_nat_2011_CISS, Xie_2011_CISS, O.Ben_nat_2017_CISS, K.Michaeli_PNAS_2019_CISS, R.Neeman_ACR_2020_CISS, D.H.Waldeck2021aplmat_CISS, Nakajima23} to inorganic metals~\cite{A.Inui2020prl_CrNb3S6, Y.Nabei_APL_2020_CrNb3S6, K.Shiota2021prl_disilicide_CISS, H.Shishido_apl_2021_NbSi2-TaSi2}. 

\begin{figure}[t!]
  \centering
  \includegraphics[width=\linewidth]{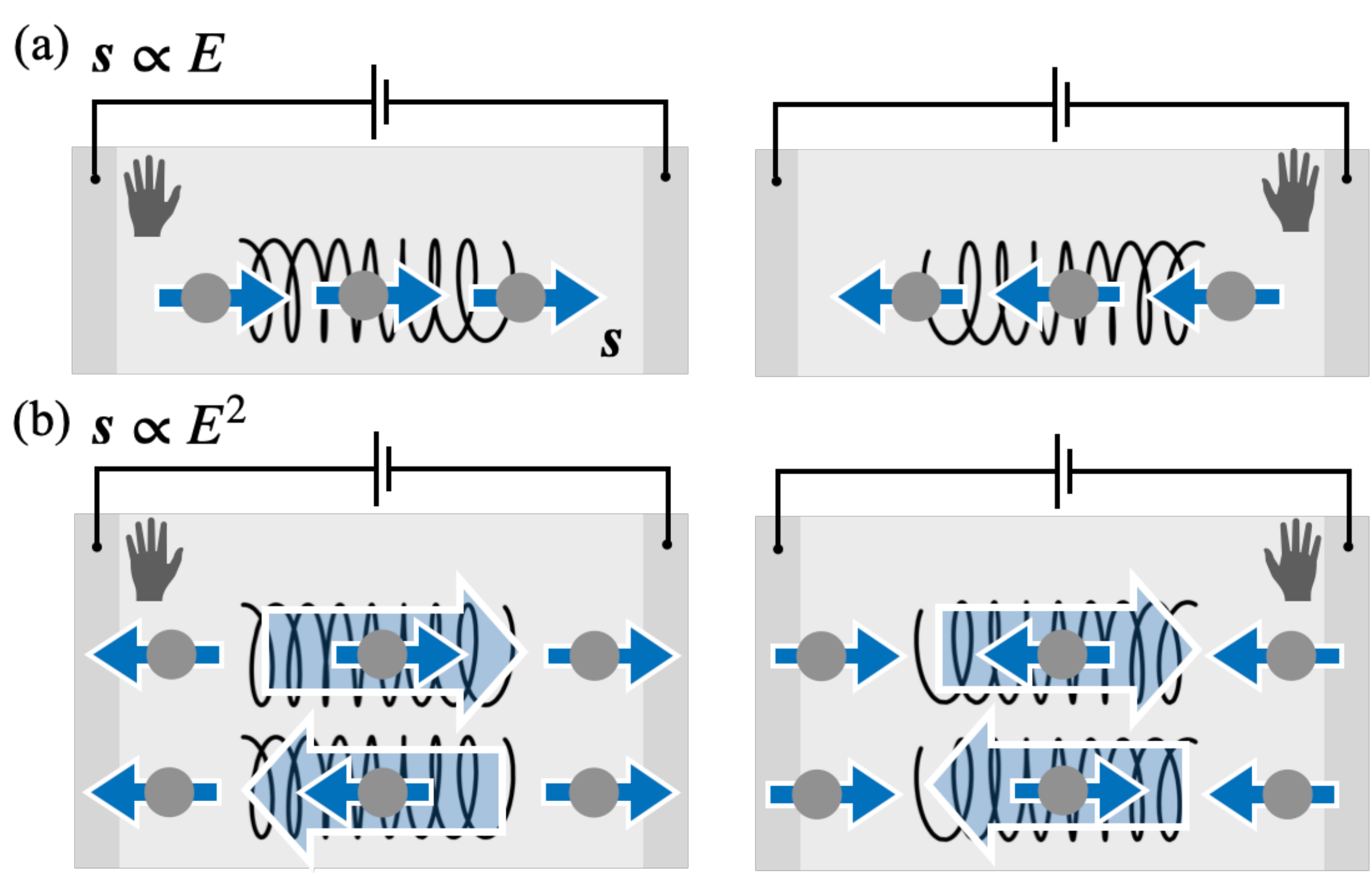}
  \caption{
  \label{f:linear_quadratic}
  Schematic of chirality-dependent spin $\bm{s}$ generation as (a) linear and (b) quadratic responses to an electric field $\bm{E}$. 
  Blue and light-blue arrows represent spins and the spin-current flow, respectively. 
  See also in \cite{yoshimi2025}.
  }
\end{figure}

In particular, CISS experiments in inorganic metals have observed induced magnetization along the current direction as a linear response to the applied electric field~\cite{A.Inui2020prl_CrNb3S6, Y.Nabei_APL_2020_CrNb3S6, K.Shiota2021prl_disilicide_CISS, H.Shishido_apl_2021_NbSi2-TaSi2}.
This is analogous to the Edelstein effect~\cite{edelstein1990spin, T.Yoda_sr_2015_Edelstein, T.Furukawa_prr_2021_Edelstein, Furukawa2017} as shown in Fig.~\ref{f:linear_quadratic}(a), where the net magnetization occurs depending on the handedness. 

On the other hand, the generation of antiparallel spin polarizations, whose spin orientation depends on the handedness, is argued as a quadratic response to the electric field, as shown in Fig.~\ref{f:linear_quadratic}(b). 
Indeed, the experimental indications on antiparallel spin polarizations have been reported in chiral organic superconductors~\cite{Nakajima23} and in enantioselective adsorption on ferromagnetic substrates~\cite{K.Banerjee-Ghosh_science_2018_Separation_of_enantiomers, Safari_admat_2024_enantioselective_adsorption, S.Miwa_sciadv_2025_spin-polarization_enantioselectivity}. 

Motivated by the pronounced contrast between the first- and second-order responses, Yoshimi and his co-workers investigated the nonlinear electric-field response of a system with explicit boundaries designed to closely replicate the experimental conditions~\cite{yoshimi2025}.
By using the Boltzmann framework with the proper constraint of Gauss law, they revealed that both bulk spin polarization in the linear response and antiparallel spin polarization near the boundary in the quadratic response in the chirality dependent manner.
In particular, generated local electric field near boundaries plays an important role to generate the antiparallel spin polarization in quadratic response.
This observation raises questions about the naive picture of spin polarizations driven by spin-current inflow.

In this paper, we also theoretically investigate the spatial distribution of the antiparallel spin polarizations generated as a quadratic response in a finite chiral system composed of triangular prisms with the ETM degree of freedom. 
By using both the nonlinear Kubo formalism and real-time numerical simulations, we show that a \textit{homogeneous} AC electric field induces an \textit{inhomogeneous} DC spin generation accompanied by dissipation. 
We further clarify that, for the present model, the handedness of chirality encoded in these chirality-dependent antiparallel spin polarizations is characterized by the sign of the spin-dependent hopping along the triangular-prism axis.
The present results are complementary to those obtained within the Boltzmann framework, in that our approach explicitly incorporates interband processes and allows access to strongly dissipative and nonpertubative regimes in numerical simulations, although the strict enforcement of Gauss law at the surfaces is not included.

The rest of this paper is organized as follows. 
In Sec.~\ref{sec:model}, we introduce the model Hamiltonian describing a finite triangular-prism system with ETMs and show the equilibrium properties. 
Then, we evaluate the quadratic antiparallel spin polarizations generation based on the Kubo formalism in Sec.~\ref{sec:quadratic_spin}. 
We further find that the naive relation between the generation rate of the spin polarization and the spatial gradient of the ETM, conventionally regarded as the spin current, is not generally valid, indicating that the source contribution in the spin continuity equation is important.
In addition, we show the contribution of ETMs to the chirality-dependent antiparallel spin polarizations generation in Sec.~\ref{sec:role_ETM}. 
Based on these results, we perform real-time simulations to investigate the spin polarization beyond the perturbative regime in Sec.~\ref{sec:real-time}.
Finally, we summarize our results in Sec.~\ref{sec:summary}.

\section{Model}
\label{sec:model}
\begin{figure}[t!]
  \centering
  \includegraphics[width=\linewidth]{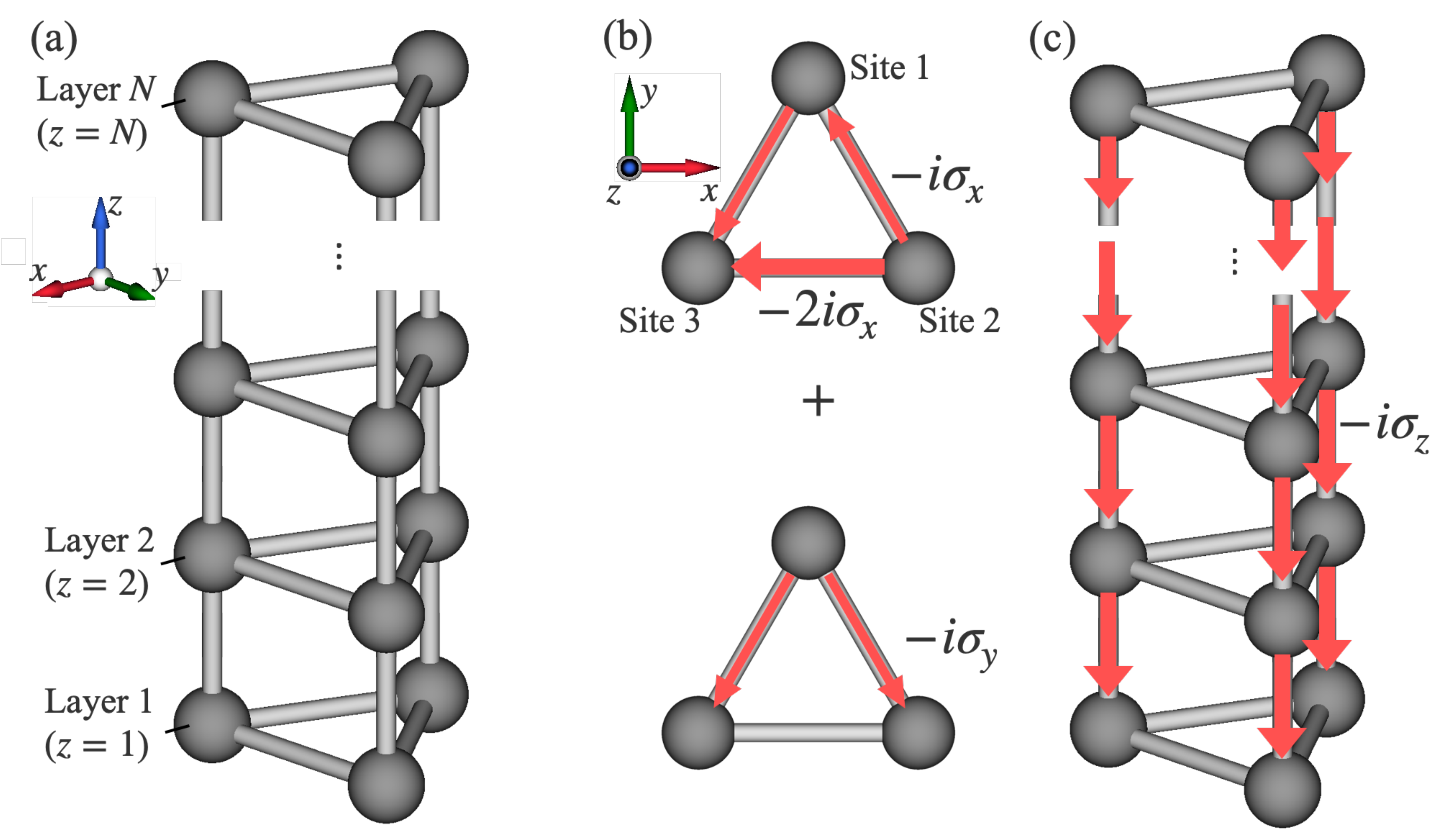}
  \caption{
  \label{f:model}
  (a) Structure of the finite $N$-layer triangular-prism system along the $z$ axis under the point group $D_{\rm 3h}$.
  (b) In-plane and (c) out-of-plane ETM degrees of freedom belonging to the $A_1''$ irreducible representation in $D_{\rm 3h}$.
  Red arrows represent imaginary hoppings with polar property and $\sigma_\nu$ for $\nu=x,y,z$ denote the Pauli matrix in spin space. 
  }
\end{figure} 

We start by considering a finite triangular-prism system along the $z$-axis under the achiral $D_{\rm 3h}$ symmetry. 
The system consists of $3N$ sites with three sites in each layer and $N$ layers, as shown in Fig.~\ref{f:model}(a); the $z$-coordinate for layer $l$ is given by $z=l$.
Taking into account the $s$-orbital degree of freedom at each site, the model Hamiltonian is given by
\begin{align}
  \mathcal{H}_{\rm achial} &= \mathcal{H}^{\rm Re}_{\perp}+ \mathcal{H}^{\rm Re}_{\parallel},\\
  \mathcal{H}^{\rm Re}_{\perp} &= -t'_{\perp} \sum_{l=1}^N \sum_{\langle i,j \rangle}\sum_{\sigma} c_{i,l,\sigma}^\dagger c_{j,l,\sigma} + \text{H.c.},\\
  \mathcal{H}^{\rm Re}_{\parallel} &= -t_{\parallel} \sum_{l=1}^N \sum_{i=1,2,3}\sum_{\sigma} c_{i,l,\sigma}^\dagger c_{i,l+1,\sigma} + \text{H.c.},
\end{align}
where $c_{i,l,\sigma}^\dagger$ ($c_{i,l,\sigma}$) is the creation (annihilation) operator of an electron with spin $\sigma$ at site $i$ ($i=1,2,3$) in layer $l$ ($l=1,2,\cdots, N$) and $\langle i,j \rangle$ represents the nearest neighbors in the same layer. 
We set lattice constants $a=c=1$.
$\mathcal{H}^{\rm Re}_{\perp}$ and $\mathcal{H}^{\rm Re}_{\parallel}$ represent the real hoppings in the $xy$ plane and along the $z$ axis, respectively.

Under the achiral $D_{3\rm h}$ symmetry, the system has mirror symmetries $\sigma_{\rm h}$ and $\sigma_{\rm v}$.
When these mirror symmetries are lost, the system becomes chiral. 
This situation can be realized by introducing the spin-dependent imaginary hopping term, which is described by
\begin{align}
  \mathcal{H}^{\rm Im}_{\perp} &= -t''_{\perp} \sum_{l=1}^{N} \vec{c}^{\,\dagger}_{l} G_{0\perp} \vec{c}_{l}, \\
  \mathcal{H}^{\rm Im}_{\parallel} &= -t''_{\parallel} \sum_{l=1}^{N-1} \vec{c}^{\,\dagger}_{l} G_{0\parallel}
  \vec{c}_{l+1} + {\rm H.c.},
\end{align}
where $\vec{c}^{\,\dagger}_l= $ $(c^\dagger_{1,l,\uparrow},$ $c^\dagger_{1,l,\downarrow},$ $c^\dagger_{2,l,\uparrow},$ $c^\dagger_{2,l,\downarrow},$ $c^\dagger_{3,l,\uparrow},$ $c^\dagger_{3,l,\downarrow})$, and the matrix representations of $G_{0\perp}$ and $G_{0\parallel}$ are given in Appendix~\ref{appendix:Hamiltonian}. 
$\mathcal{H}_{{\perp}}^{\rm Im}$ and $\mathcal{H}_\parallel^{\rm Im}$ denote the $xy$-plane and $z$-axis contributions, respectively, both of which belong to the $A''_1$ representation under $D_{\rm 3h}$, corresponding to the ETM degree of freedom~\cite{hayami2018prb_Classification_of_atomic-scale_multipoles}. 
The schematics of the spin-dependent imaginary hopping processes are presented in Figs.~\ref{f:model}(b) and (c), which can be regarded as the microscopic $\nu$-polarized ``spin current'' along the $\nu$ direction ($\nu=x,y,z$).  
By taking into account these two imaginary hopping terms, the symmetry of the system reduces to the chiral $D_3$ symmetry. 
Consequently, we analyze the Hamiltonian written as 
\begin{align}
  \mathcal{H}_0 &= \mathcal{H}_{\rm achial} + \mathcal{H}^{\rm Im}_{\perp} + \mathcal{H}^{\rm Im}_{\parallel}.
\end{align}

\begin{figure}[t!]
  \centering
  \includegraphics[width=\linewidth]{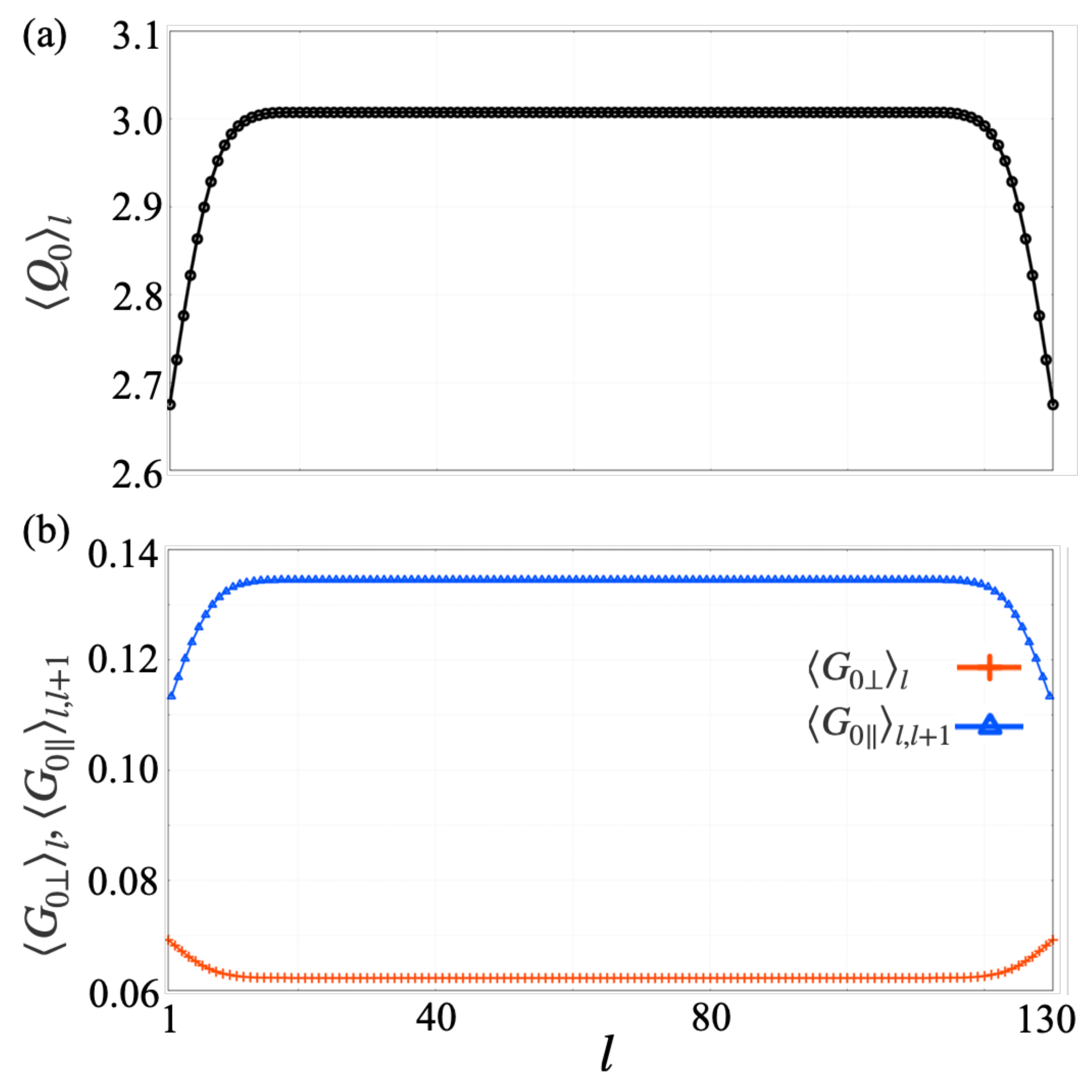}
  \caption{
  \label{f:Q0_G0xy_G0z}
  (a) Layer dependence of the electric monopole (onsite potential) $\braket{Q_{0}}_{l}$ and (b) the ETMs (spin-dependent imaginary hoppings) $\braket{G_{0\perp}}_{l}$ and $\braket{G_{0\parallel}}_{l, l+1}$ for the equilibrium state. The model parameters are $t'_{\perp}=2$, $t'_{\parallel}=1$, $t''_{\perp}=0.5$, $t''_{\parallel}=0.3$, and $T=0.01$. 
}
\end{figure}

As preliminary analysis, the spatial distributions of ETMs, as well as the charge potentials, are evaluated by diagonalizing the total Hamiltonian given in the above. 
Figures~\ref{f:Q0_G0xy_G0z}(a) and (b) represent the expectation values of the layer dependence of the electric monopole for a each layer $l$, $\langle Q_{0}\rangle_l$, corresponding to the onsite potential (the explicit form is shown in Appendix~\ref{appendix:Hamiltonian}) and in-plane and out-of-plane ETMs, $\braket{G_{0\perp}}_{l}$ and $\braket{G_{0\parallel}}_{l,l+1}$, for the equilibrium state; $\langle \cdots \rangle_l$ and $\braket{\cdots}_{l,l+1}$ denote the expectation values for $l$ layer and $l$-$l+1$ inter layer, respectively. 
The model parameters are set to $N=130$, $t'_{\perp}=2$, $t'_{\parallel}=1$, $t''_{\perp}=0.5$, $t''_{\parallel}=0.3$, and the temperature $T=0.01$. 
The chemical potential $\mu=0.5838443$ lies at the midpoint between the $(3N)$-th and $(3N+1)$-th energy levels, counted from the lowest, corresponding to the half-filling condition at zero temperature. 
Here and hereafter, the layer dependence of the physical quantity $\Lambda$ is smoothed by applying a Gaussian filter with a width of $\sigma = 5$ layers under the Neumann boundary condition $\partial_z \Lambda|_{z=0, N} = 0$.

As shown in Fig.~\ref{f:Q0_G0xy_G0z}(a), $\braket{Q_{0}}_{l}$ is nearly constant in the layers near the center, and it is deviated from the averaged value near the surface layers. 
This indicates that the surface polarization along the $z$ axis is induced antiparallelly owing to the edge under the open boundary. 
Similarly, both $\braket{G_{0\perp}}_{l}$ and $\braket{G_{0\parallel}}_{l, l+1}$ are finite throughout the system, as shown in Fig.~\ref{f:Q0_G0xy_G0z}(b), which is attributed to nonzero $t''_{\perp}$ and $t''_{\parallel}$. 
The fact that the ETMs have homogenous sign thgought layers indicates that the present triangular-prism system forms an enantiopure cluster.

\section{Quadratic antiparallel spin polarization and electric-toroidal monopole} 
\label{sec:quadratic_spin}

\begin{figure}[t!]
  \centering
  \includegraphics[width=\linewidth]{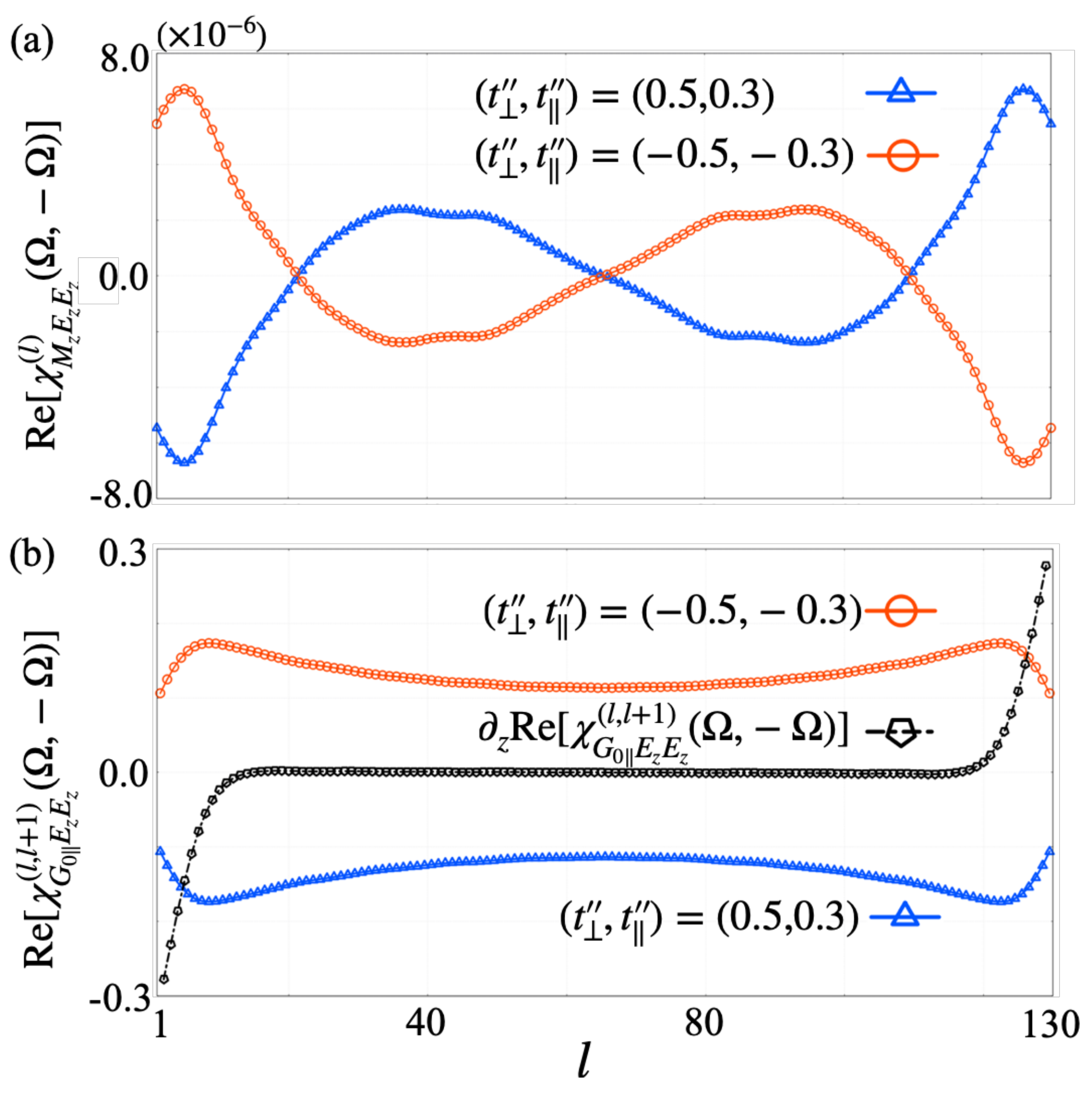}
  \caption{
  \label{f:chi_G0zEzEz_chi_szEzEz}
  Layer dependence of (a) $\chi^{(l)}_{M_zE_zE_z}(\Omega, -\Omega)$  and (b) $\chi^{(l,l+1)}_{G_{0\parallel}E_zE_z}(\Omega, -\Omega)$ for $\Omega = 1$.
  The other parameters are the same as those of Fig.~\ref{f:Q0_G0xy_G0z}, and the handedness of the blue and red lines is opposite to each other.
  Black line in (b) represents $\partial_z \chi^{(l,l+1)}_{G_{0\parallel} E_zE_z}(\Omega, -\Omega)$ for $(t''_{\perp}, t''_{\parallel})= (0.5, 0.3)$.
}
\end{figure}

Next, let us investigate the chirality-dependent antiparallel spin polarizations induced by an AC electric field in the triangular-prism system.
For that purpose, we evaluate the layer-resolved magnetization under the AC electric field.
By focusing on the DC magnetization and longitudinal responses, the lowest-order contribution from the electric field arises at second order, where the applied field consists of the photocurrent-type components at frequencies $+\Omega$ and $-\Omega$.
This is given by 
\begin{align}
  M^{(l)}_{z}(\omega=0) 
  &=  
  \chi^{(l)}_{M_zE_zE_z}(\Omega, -\Omega) E_z(\Omega) E_z(-\Omega),
\end{align}
where we consider the $z$-directional response along the chiral axis.
Here, $M^{(l)}_{z}$ denotes the total $z$-component of the magnetization obtained by summing over the three-site spin moments in the $xy$ plane of layer $l$.

The quadratic longitudinal magneto-electric tensor along $z$ axis $ \chi^{(l)}_{M_zE_zE_z}(\Omega, -\Omega)$ is calculated by the standard Kubo formalism as follows: 
\begin{align}
  &\chi^{(l)}_{M_zE_zE_z}(\Omega, -\Omega)\notag\\ 
  &= \frac{1}{2}\sum_{ijk}\Bigg[ \frac{s_{z,l}^{ij}}{2i\hbar\delta + (\xi_i - \xi_j)}  \left\{ \frac{f(\xi_i) - f(\xi_k)}{\hbar\Omega + i\hbar\delta + (\xi_i - \xi_k)} q_{z,l}^{jk} q_{z,l}^{ki} \right. \notag\\
  & \left. \quad - \frac{f(\xi_k) - f(\xi_j)}{\hbar\Omega + i\hbar\delta + (\xi_k - \xi_j)} q_{z,l}^{jk} q_{z,l}^{ki} \right\}  + (\Omega\rightarrow -\Omega)\Bigg],\label{eq:chi_szEzEz}
\end{align}
where $\xi_i$ is the $i$th energy eigenvalues, $f(\xi_i)$ is the Fermi-Dirac distribution function, $s_{z,l}^{ij}= \langle i |s_{z}| j \rangle_l$ and $q_{z,l}^{ij}=\braket{i|qz|j}_l$ are the matrix elements of the $l$th-layer spin operator $s_z$ and the electric polarization operator $ql$ ($q$ is the elementary charge) between the eigenstates $|i\rangle$ and $|j\rangle$, respectively. 
$\hbar$ is the reduced Planck constant. 
$\delta = 1/\tau$ is the scattering rate represented as the inverse of the relaxation time $\tau$, and we take $\delta=0.01$. 
Hereafter, we set $\hbar = q = 1$.
Note that the electric polarization has a constant ambiguity depending on the choice of the reference point.

From the symmetry view point, such a second-order DC spin (magnetization) response to the electric field is allowed only when accompanied by a spatial modulation per layer; no uniform magnetization $M_z=\sum_l M^{(l)}_{z}=0$ when the twofold rotational symmetry perpendicular to the principal axis remains~\cite{kirikoshi2025light}. 
Specifically, an antiparallel-type magnetization defined with respect to the center along the prism axis is symmetry-allowed in chiral systems, i.e., $M^{(l)}_{z}=-M^{(N-l+1)}_{z}$ (or $\chi^{(l)}_{M_zE_zE_z}(\Omega, -\Omega)=-\chi^{(N-l+1)}_{M_zE_zE_z}(\Omega, -\Omega)$), which corresponds to the antiparallel spin polarizations, as shown in Fig.~\ref{f:linear_quadratic}(b).  
This response is also understood by the magnetic monopole response against the electric field, since the antiparallel spin magnetization can be associated with the magnetic monopole defined as $M^{(l)}_0 \equiv (M^{(l)}_{z}-M^{(N-l+1)}_{z})/2$. 
Owing to the presence of the time-reversal symmetry, such a magnetic monopole response, $M^{(l)}_0 \propto E_z^2$, requires dissipative effects ($\delta \neq 0$)~\cite{yatsushiro2021prb_122}. 
It is noted that the linear-order DC spin response to the AC electric field is prohibited by the symmetry, and the induced magnetization oscillates in phase with the frequency of the applied electric field.
On the contrary, the DC response is allowed in the quadratic response. Therefore, applying the AC electric field has an advantage for detecting the DC antiparallel spin polarizations.

Figure~\ref{f:chi_G0zEzEz_chi_szEzEz}(a) shows the layer dependence of the quadratic response tensor $\chi^{(l)}_{M_zE_zE_z}(\Omega, -\Omega)$ for $\Omega=1$ and $N=130$. 
The red (blue) symbols represent the data for opposite handedness by setting $t''_{\perp}=0.5$ and $t''_{\parallel}=0.3$ ($t''_{\perp}=-0.5$ and $t''_{\parallel}=-0.3$); the other parameters are the same as Sec.~\ref{sec:model}.
The opposite signs of $\chi^{(l)}_{M_zE_zE_z}(\Omega, -\Omega)$ against the opposite handedness indicates that the origin of this quadratic response lies in the intrinsic chirality of the system; indeed, $\chi^{(l)}_{M_zE_zE_z}(\Omega, -\Omega)=0$ in the achiral system with $t''_{\perp}=t''_{\parallel}=0$.
In addition, the relation of $\chi^{(l)}_{M_zE_zE_z}(\Omega, -\Omega)=\chi^{(N-l+1)}_{M_zE_zE_z}(\Omega, -\Omega)$ holds while the total contribution $\sum_l \chi^{(l)}_{M_zE_zE_z}(\Omega, -\Omega)$ vanishes, which is consistent with a magnetic-monopole--type response, as described above.
Thus, the external AC electric field induces DC antiparallel spin polarizations in a chirality-dependent manner through its quadratic response. 

In addition, we investigate the relaxation-time $\tau$ dependence of $\chi^{(l)}_{M_zE_zE_z}(\Omega, -\Omega)$ for the small number of layer $N=4$ and $\Omega=1.00064$, which corresponds to the resonance frequency (not shown). 
We find that $\chi^{(l)}_{M_zE_zE_z}(\Omega, -\Omega)$ is always proportional to the odd power of $\tau$, which is consistent with the time-reversal property of the tensor $\chi^{(l)}_{M_zE_zE_z}$ defined in Eq.~(\ref{eq:chi_szEzEz}).

The above results imply that the spin seems to accumulate in each layer when the AC electric field is applied.
In other words, such spin accumulation is customarily interpreted as a consequence of spin-current through the phenomenological spin continuity equation, we examine their relationship. 
To this end, we evaluate the second-order spin current response against the electric field. 
In this case, as the ETM corresponds to the ``spin current", we evaluate the quadratic response $\chi^{(l, l+1)}_{G_{0\parallel}E_zE_z}(\Omega, -\Omega)$, by using the Kubo formula with the replacement of $s_{z,l}^{ij}$ with $G_{0\parallel, l,l+1}^{ij}$ in Eq.~(\ref{eq:chi_szEzEz}). 
The data are plotted in Fig.~\ref{f:chi_G0zEzEz_chi_szEzEz}(b), where the handedness-dependent response is obtained, similar to the spin response in Fig.~\ref{f:chi_G0zEzEz_chi_szEzEz}(a). 
By supposing that the generating spin current contributes to the spin accumulation, its spatial gradient of $\chi^{(l,l+1)}_{G_{0\parallel}E_zE_z}(\Omega, -\Omega)$, i.e, $\partial_z \chi^{(l,l+1)}_{G_{0\parallel}E_zE_z}(\Omega, -\Omega)$, should be related to $\chi^{(l)}_{M_zE_zE_z}(\Omega, -\Omega)$.  
However, as shown by the black line in Fig.~\ref{f:chi_G0zEzEz_chi_szEzEz}(b), the behavior of $\partial_z \chi^{(l,l+1)}_{G_{0\parallel}E_zE_z}(\Omega, -\Omega)$ seems to be uncorrelated to $\chi^{(l)}_{M_zE_zE_z}(\Omega, -\Omega)$, which indicates that the naive relation between the ETM (spin current) and spin polarizations does not hold and the significance of the source-term contribution.
This discrepancy is qualitatively consistent with the observation reported in Ref.~\cite{yoshimi2025} where the prominence of source-field contributions is attributed to local electric fields arising from the inhomogeneous potential near the surfaces [see Fig.~\ref{f:Q0_G0xy_G0z}(a)].

\section{Role of electric toroidal monopole}
\label{sec:role_ETM}

\begin{figure}[t!]
  \centering
  \includegraphics[width=\linewidth]{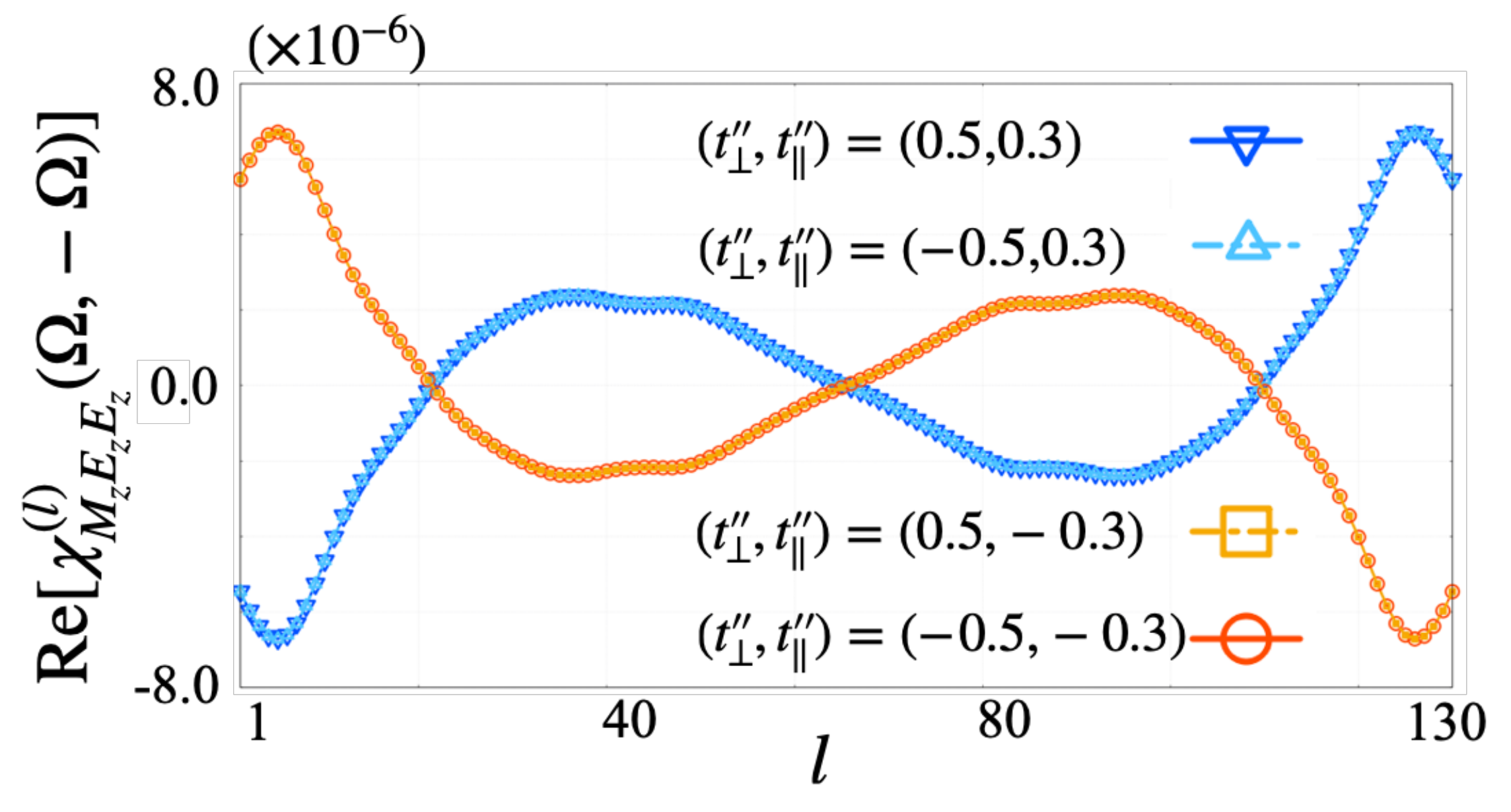}
  \caption{
  \label{f:opposite-tG0}
  Layer dependence of $\chi^{(l)}_{M_zE_zE_z}(\Omega, -\Omega)$ for $(t''_{\perp}, t''_{\parallel})=(0.5, 0.3)$, $(0.5, -0.3)$, $(-0.5, 0.3)$, and $(-0.5, -0.3)$. The other parameters are the same as those of Fig.~\ref{f:chi_G0zEzEz_chi_szEzEz}.
  }
\end{figure}
\begin{figure}[t!]
  \centering
  \includegraphics[width=\linewidth]{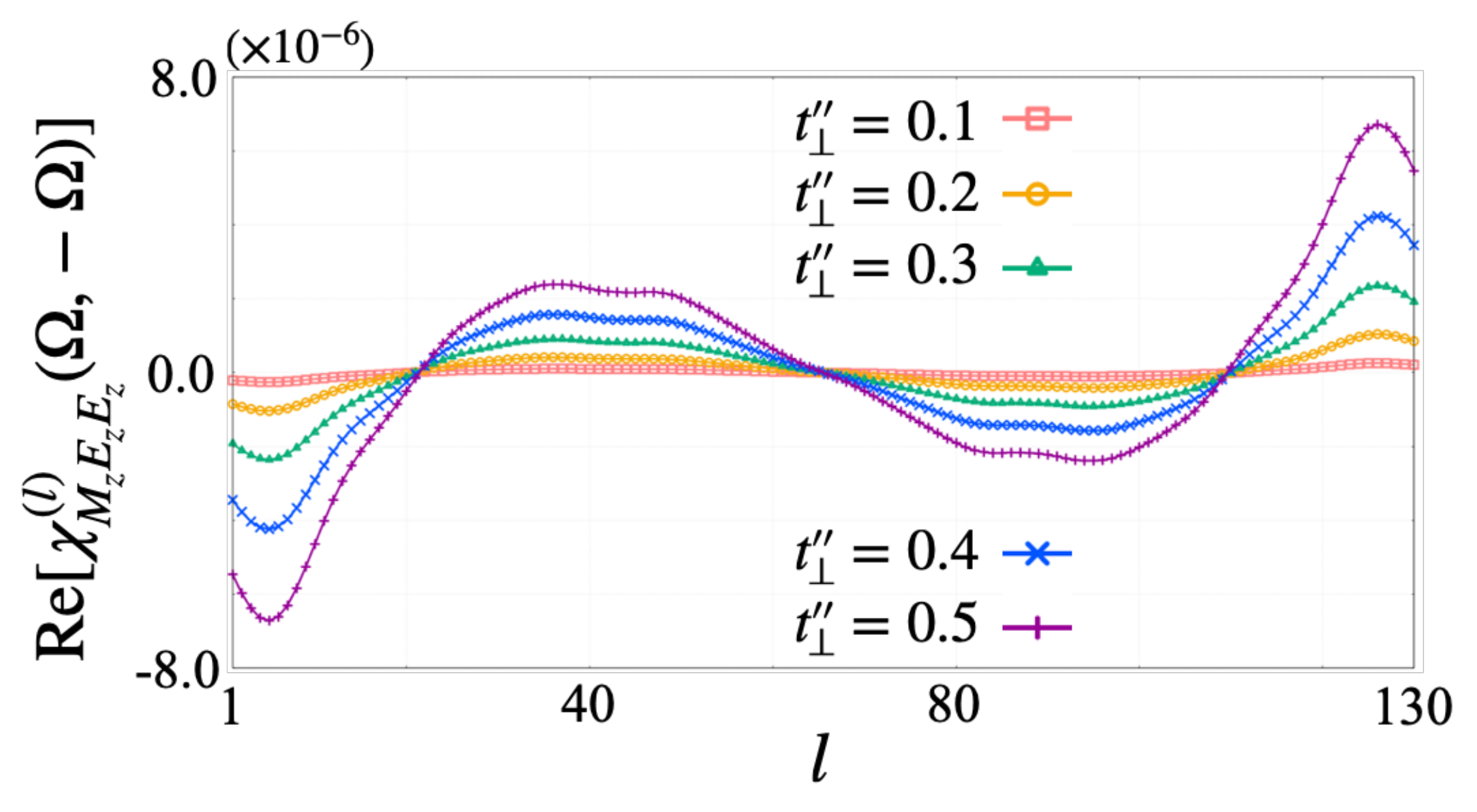}
  \caption{
  \label{f:tG0xy-dep}
  The in-plane ETM $t''_{\perp}$ dependence of $\chi^{(l)}_{M_zE_zE_z}(\Omega, -\Omega)$. 
  The other parameters are the same as those of Fig.~\ref{f:chi_G0zEzEz_chi_szEzEz}. 
  }
\end{figure}

Since the induced antiparallel spin polarizations is chirality(handedness)-dependent, its microscopic origin is related to the symmetry-breaking hoppings, $t''_{\perp}$ and $t''_{\parallel}$, which correspond to the in-plane and out-of-plane ETMs, respectively. 
It should be emphasized that the handedness is reversed when \textit{both} of $t''_{\perp}$ and $t''_{\parallel}$ change signs. Now, we show the relationship between $(t''_{\perp}, t''_{\parallel})$ and $\chi^{(l)}_{M_zE_zE_z}(\Omega, -\Omega)$ in Fig.~\ref{f:opposite-tG0} by changing the signs of the spin-dependent hoppings. 
The sign of $\chi^{(l)}_{M_zE_zE_z}(\Omega, -\Omega)$ is reversed when the sign of $t''_{\parallel}$ is reversed, whereas it remains unchanged when the sign of $t''_{\perp}$ is reversed. 
These behaviors indicate that the out-of-plane ETM $G_{0\parallel}$ determines the handedness property in terms of the chirality-dependent antiparallel spin polarizations. 
Meanwhile, the in-plane ETM $t''_{\perp}$ plays an important role in enhancing $\chi^{(l)}_{M_zE_zE_z}(\Omega, -\Omega)$; the response vanishes at $t''_{\perp}=0$, and is developed with the increase of $t''_{\perp}$, as shown in Fig.~\ref{f:tG0xy-dep}.

For further insight into these behaviors, we analyze the essential model parameters responsible for inducing the chirality-dependent antiparallel spin polarizations. 
Performing the expansion method for $\chi^{(l)}_{M_zE_zE_z}(\Omega, -\Omega)$~\cite{hayami2020prb_bottom-up, R.Oiwa2022jpsj_nonl-respo}, we find that the model-dependent part of $\chi^{(l)}_{M_zE_zE_z}(\Omega, -\Omega)$ always includes the factor $t'_{\parallel} {t''_{\perp}}^2 t''_{\parallel}$ in the expansion. 
This result supports our numerical findings that the sign of $\chi^{(l)}_{M_zE_zE_z}(\Omega, -\Omega)$ is governed by $t''_{\parallel}$, rather than by $t''_{\perp}$. 
On the other hand, we confirm that the essential model parameters for the longitudinal second-order magneto-electric response in the in-plane direction is always proportional to the odd power of $t_{\perp}''$ and even power of $t_{\parallel}''$, indicating that the sign of $t_\perp''$ determines the sign of the response. 
These considerations highlight the importance of the two ETM degrees of freedom for the chirality-dependent antiparallel spin polarizations: in particular, for antiparallel spin polarizations along $z$ axis, the out-of-plane ETM determines its sign, whereas the in-plane ETM modulates its magnitude.
In addition, it is noted that the spin-independent hopping along the $z$ direction, $t'_{\parallel}$, is also required in the present system, which indicates that only the spin-dependent hopping is not enough to induce the antiparallel spin polarizations, which implies that the (local) spin polarization occurs through the real (local) hopping process along $z$ direction. i.e., (local) Edelstein effect.

\section{Spin polarizations beyond perturbative regime}
\label{sec:real-time}

Finally, we perform real-time simulations to evaluate the spin polarization beyond the perturbative regime within the Kubo formalism.
We directly solve the time-dependent Hamiltonian, which is given by
\begin{align}
  \mathcal{H}(t) &= \mathcal{H}^{\rm Re}_{\perp}+ \mathcal{H}^{\rm Im}_{\perp}
  + \tilde{\mathcal{H}}^{\rm Re}_{\parallel}(t) + \tilde{\mathcal{H}}^{\rm Im}_{\parallel}(t),\\
  \tilde{\mathcal{H}}^{\rm Re}_{\parallel}(t) 
  &=
  -t'_{\parallel}e^{-i q A_z(t)} \sum_{l=1}^N \sum_{i=1,2,3}\sum_{\sigma} c_{i,l,\sigma}^\dagger c_{i,l+1,\sigma}
  \\
  &+ \text{H.c.},\\
  \tilde{\mathcal{H}}^{\rm Im}_{\parallel}(t) 
  &= 
  -t''_{\parallel}e^{-i q A_z(t)}  \sum_{l=1}^{N-1} c^\dagger_{l} G_{0\parallel}c_{l+1}+ {\rm H.c.},
\end{align}
where we adopt the Peierls substitution to introduce the electric field under the velocity gauge, which modulates the phase factor of the hoppings along the electric-field direction, $\mathcal{H}^{\rm Re}_{\parallel}$ and $\mathcal{H}^{\rm Im}_{\parallel}$; the time-dependent electric field $E_z (t)$ is related to the vector potential $A_z(t)$ as $E_z(t)= -d A_z(t)/dt$.
We apply an AC electric field of the form $E_z(t) = E_0 \cos(\Omega t)$, for which the corresponding vector potential is given by $A_z(t) = - (E_0/\Omega) \sin(\Omega t)$.

The time-evolution of the system is described by the single-particle density matrix $\rho_{ij}(t)=\langle c_i^{\dag}c_j\rangle (t)$ $\equiv {\rm Tr}[c_i^\dagger c_j\rho(t)]$, where $\rho(t)$ is the density matrix of the system at time $t$, which obeys the following von Neumann equation~\cite{Yue:22}:
\begin{align}
  \frac{d \rho(t)}{dt} &= -i [\mathcal{H}(t), \rho(t)] - \gamma \{\rho(t) - \rho_{\rm eq}(t)\}, \label{eq:vonNeumann}
\end{align}
where $\rho_{\rm eq}(t)$ is the single-particle density matrix in equilibrium: $\rho_{\rm eq}(t) = \sum_i f(\xi_i(t)) |\psi_i(t)\rangle \langle \psi_i(t)|$, where $\mathcal{H}(t)|\psi_i(t)\rangle = \xi_i(t) |\psi_i(t)\rangle$.
We incorporate the relaxation effect phenomenologically via the relaxation time approximation in the second term of Eq.~(\ref{eq:vonNeumann})~\cite{PhysRevB.110.125111,hattori2024effect,xsrf-t1hj,qsxr-c2pq,PhysRevB.106.035204}. 
The relaxation time is assumed to be isotropic and spin-independent, and is parameterized as $\gamma = 1/\tau$; the relaxation time $\tau=1/\delta$ corresponds to that in the Kubo formalism in Eq.~(\ref{eq:chi_szEzEz}). 
We compute the time evolution of $\rho(t)$ by using the fourth-order Runge--Kutta method and evaluate the magnetization as $\langle M_{z}^{(l)}\rangle (t) = {\rm Tr}[\rho(t) s_{z}^{(l)}]$. 
To evaluate the static component of the spin polarizations $\langle M_z^{(l)}(\omega=0)\rangle$, following the previous studies~\cite{PhysRevLett.127.127402, PhysRevB.109.064407, hattori2025nonlinear, hattori2025dirac}, we perform the Fourier transformation of $
\langle M_{z}^{(l)}\rangle (t)$ for $t\geq 628.32$ after the nonequilibrium steady state is obtained.

\begin{figure}[t!]
  \centering
  \includegraphics[width=\linewidth]{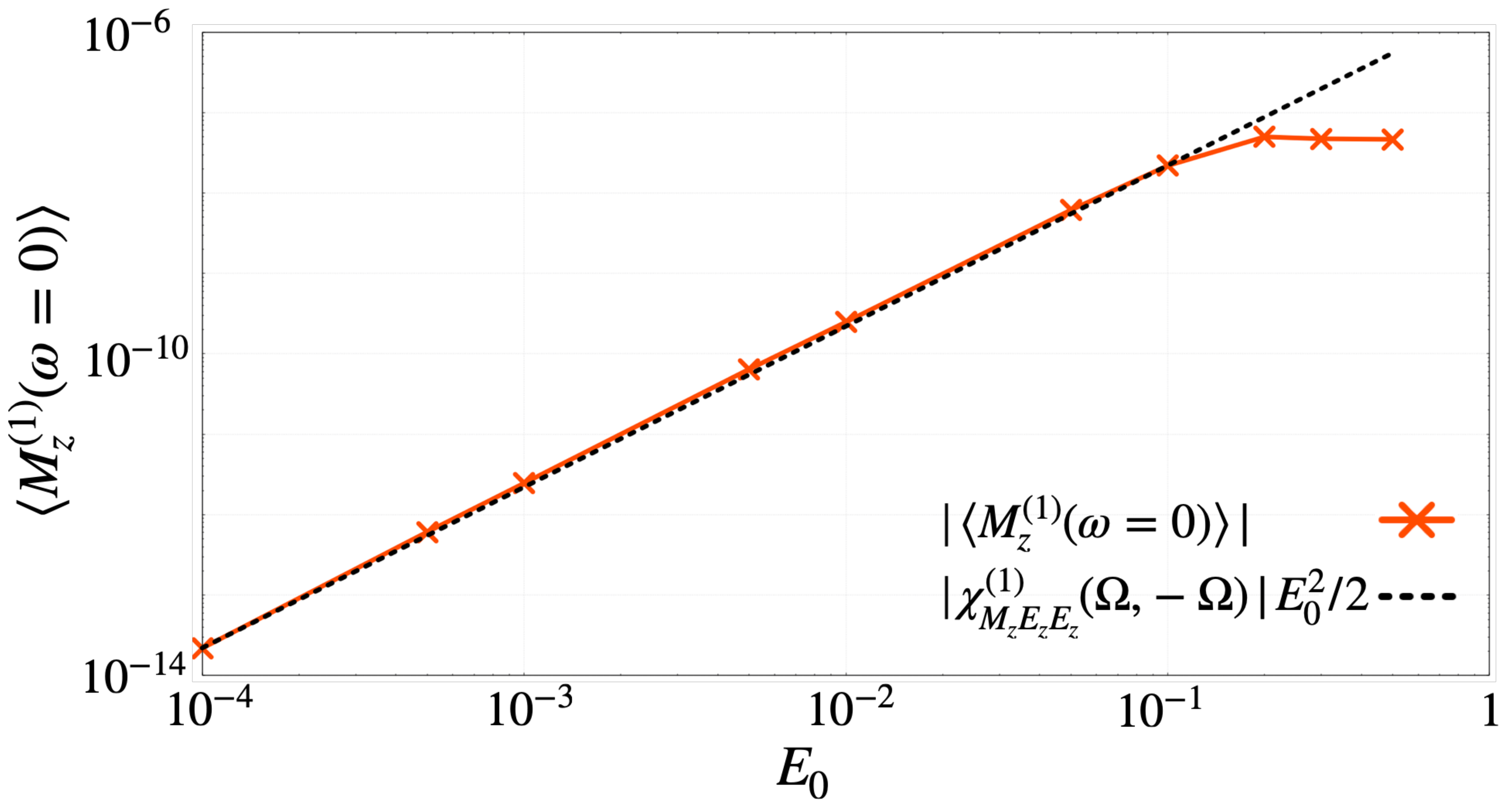}
  \caption{
  \label{f:E0-dep} 
  $E_0$ dependence of $\langle M^{(1)}_z(\omega=0)\rangle$ at $\Omega = 1$ and $N=40$.
  The other parameters are the same as those of Fig.~\ref{f:chi_G0zEzEz_chi_szEzEz}.
  The solid orange line represents $|\chi^{(1)}_{M_zE_zE_z}(\Omega, -\Omega)| E_0^2/2$, which is evaluated by the Kubo formula with the same size, $N=40$ in Eq.~(\ref{eq:chi_szEzEz}).
  }
\end{figure}

Figure~\ref{f:E0-dep} shows the $E_0$ dependence of the DC spin magnetization at the edge, $\langle M^{(1)}_{z}(\omega=0)\rangle = M^{(1)}_0$, which is obtained by soliving Eq.~(\ref{eq:vonNeumann}) for $\Omega=1$, $\delta=0.01$, and $N=40$.
We also plot the results in the Kubo formalism for comparison.
When the amplitude of the electric field is small ($E_0 \lesssim 10^{-1}$), the results of the time-dependent simulations are scaled as $E_0^2$, which agree well with those obtained from the Kubo formalism.
The simulation results deviates from the $E_0^2$ scaling around $E_0\sim 10^{-1}$ as higher-order contributions become significant. 
This result indicates that a large electric field facilitates the observation of antiparallel spin polarizations; however, if the field strength becomes too large, their magnitude eventually saturates.

\section{Summary} 
\label{sec:summary}
We have investigated the chirality-dependent antiparallel spin polarization generation in a finite triangular-prism system hosting ETMs, which serve as an electronic order parameter of chirality. 
Based on both the nonlinear Kubo formula and real-time simulations, we have shown that spatially inhomogeneous DC antiparallel spin polarization are generated as the quadratic response to the homogeneous AC electric field. 
The generated spin polarization is antisymmetric with respect to the center of the system due to the overall symmetry. 
We have also demonstrated that the resulting spatial distribution of the induced antiparallel spin polarization is not correlated to the spin-current (ETM) distribution indicating the significance of the source-field contribution due to the electric induction. 
Furthermore, we have revealed that the out-of-plane ETM determines the sign of the chirality-dependent magneto-electric responses along $z$ direction, while the in-plane ETM modulates the magnitude of the response with keeping its sign.
These results will shed light on the underlying physics of the observed antiparallel spin polarizations in chiral systems~\cite{Nakajima23} and enantioselective adsorption using ferromagnetic substrates~\cite{K.Banerjee-Ghosh_science_2018_Separation_of_enantiomers, Safari_admat_2024_enantioselective_adsorption, S.Miwa_sciadv_2025_spin-polarization_enantioselectivity}. 

\begin{acknowledgments}
  We are deeply grateful to Y. Kato, H. M. Yamamoto, and Y. Togawa for fruitful discussions.
  This research was supported by JSPS KAKENHI Grants Numbers JP22H00101, JP22H01183, JP23H04869, JP23K03288, and by JST SPRING, (JPMJSP2108) (K.H.), JST CREST (JPMJCR23O4) and JST FOREST (JPMJFR2366). 
  A.I. is financially supported as a JSPS Research Fellow. 
  K.H. was supported by the Program for Leading Graduate Schools (MERIT-WINGS).
\end{acknowledgments}
\appendix
\section{Representations of $Q_{0}$ and ETMs}
\label{appendix:Hamiltonian}
The explicit expressions of matrices $Q_{0}$ for the six $s$-wave functions $\{\phi_{1,l,\uparrow},\phi_{1,l,\downarrow},\phi_{2,l,\uparrow},\phi_{2,l,\downarrow},\phi_{3,l,\uparrow},\phi_{3,l,\downarrow}\}$, $G_{0\perp}$, and $G_{0\parallel}$ are given as follows:
\begin{widetext}
\begin{align}
  Q_{0}
  &=I_6,\\
  G_{0\perp}
  &=
  \begin{bmatrix}
    0 & 0 & 0 & \frac{1}{12} i \left(\sqrt{3}+3 i\right) & 0 & -\frac{1}{12} i \left(\sqrt{3}-3 i\right)\\
    0 & 0 & \frac{1}{12} \left(3+i \sqrt{3}\right) & 0 & \frac{1}{12} \left(3-i \sqrt{3}\right)& 0 \\
    0 & \frac{1}{12} \left(3-i \sqrt{3}\right) & 0 & 0 & 0 & -\frac{i}{2 \sqrt{3}}\\
    -\frac{1}{12} i \left(\sqrt{3}-3 i\right) & 0 & 0 & 0 & -\frac{i}{2 \sqrt{3}}& 0 \\
    0 & \frac{1}{12} \left(3+i \sqrt{3}\right) & 0 & \frac{i}{2 \sqrt{3}}& 0 & 0 \\
    \frac{1}{12} i \left(\sqrt{3}+3 i\right) & 0 & \frac{i}{2 \sqrt{3}}& 0 & 0 & 0 
  \end{bmatrix},\\
  G_{0\parallel}
  &=
  \begin{bmatrix}
    i & 0 & 0 & 0 & 0 & 0\\
    0 & -i & 0 & 0 & 0 & 0\\
    0 & 0 & i & 0 & 0 & 0\\
    0 & 0 & 0 & -i & 0 & 0\\
    0 & 0 & 0 & 0 & i & 0\\
    0 & 0 & 0 & 0 & 0 & -i
  \end{bmatrix},
\end{align}
\end{widetext}
where $I_n$ is the identity matrix of dimension $n\times n$.
\bibliographystyle{apsrev4-2}
\bibliography{main}

@article{hattori2025dirac,
  title = {Dirac Charge in Antiferromagnetic Topological Semimetals},
  author = {Hattori, Kohei and Watanabe, Hikaru and Arita, Ryotaro},
  journal = {Phys. Rev. Lett.},
  volume = {136},
  issue = {4},
  pages = {046603},
  numpages = {9},
  year = {2026},
  month = {Jan},
  publisher = {American Physical Society},
  doi = {10.1103/sfc4-8wph},
  url = {https://link.aps.org/doi/10.1103/sfc4-8wph}
}

@article{PhysRevB.110.125111,
  title = {{High harmonic generation from electrons moving in topological spin textures}},
  author = {Ono, Atsushi and Okumura, Shun and Imai, Shohei and Akagi, Yutaka},
  journal = {Phys. Rev. B},
  volume = {110},
  issue = {12},
  pages = {125111},
  numpages = {17},
  year = {2024},
  month = {Sep},
  publisher = {American Physical Society},
  doi = {10.1103/PhysRevB.110.125111},
  url = {https://link.aps.org/doi/10.1103/PhysRevB.110.125111}
}

@article{xsrf-t1hj,
  title = {{Temporal modulation of second harmonic generation in ferroelectrics by a pulsed electric field}},
  author = {Ono, Atsushi},
  journal = {Phys. Rev. B},
  volume = {112},
  issue = {11},
  pages = {115146},
  numpages = {11},
  year = {2025},
  month = {Sep},
  publisher = {American Physical Society},
  doi = {10.1103/xsrf-t1hj},
  url = {https://link.aps.org/doi/10.1103/xsrf-t1hj}
}

@article{qsxr-c2pq,
  title = {{Extracting Nonlinear Dynamical Response Functions from Time Evolution}},
  author = {Ono, Atsushi},
  journal = {Phys. Rev. Lett.},
  volume = {135},
  issue = {2},
  pages = {026401},
  numpages = {11},
  year = {2025},
  month = {Jul},
  publisher = {American Physical Society},
  doi = {10.1103/qsxr-c2pq},
  url = {https://link.aps.org/doi/10.1103/qsxr-c2pq}
}

@article{PhysRevB.106.035204,
  title = {{Doping and gap size dependence of high-harmonic generation in graphene: Importance of consistent formulation of light-matter coupling}},
  author = {Murakami, Yuta and Sch\"uler, Michael},
  journal = {Phys. Rev. B},
  volume = {106},
  issue = {3},
  pages = {035204},
  numpages = {13},
  year = {2022},
  month = {Jul},
  publisher = {American Physical Society},
  doi = {10.1103/PhysRevB.106.035204},
  url = {https://link.aps.org/doi/10.1103/PhysRevB.106.035204}
}

@article{PhysRevLett.127.127402,
  title = {{Bulk Photovoltaic Effect Driven by Collective Excitations in a Correlated Insulator}},
  author = {Kaneko, Tatsuya and Sun, Zhiyuan and Murakami, Yuta and Gole\ifmmode \check{z}\else \v{z}\fi{}, Denis and Millis, Andrew J.},
  journal = {Phys. Rev. Lett.},
  volume = {127},
  issue = {12},
  pages = {127402},
  numpages = {6},
  year = {2021},
  month = {Sep},
  publisher = {American Physical Society},
  doi = {10.1103/PhysRevLett.127.127402},
  url = {https://link.aps.org/doi/10.1103/PhysRevLett.127.127402}
}

@article{PhysRevB.109.064407,
  title = {{Bulk photovoltaic effect in antiferromagnet: Role of collective spin dynamics}},
  author = {Iguchi, Junta and Watanabe, Hikaru and Murakami, Yuta and Nomoto, Takuya and Arita, Ryotaro},
  journal = {Phys. Rev. B},
  volume = {109},
  issue = {6},
  pages = {064407},
  numpages = {17},
  year = {2024},
  month = {Feb},
  publisher = {American Physical Society},
  doi = {10.1103/PhysRevB.109.064407},
  url = {https://link.aps.org/doi/10.1103/PhysRevB.109.064407}
}

@article{Togawa2023,
	author = {Togawa ,Yoshihiko and Ovchinnikov ,Alexander S. and Kishine ,Jun-ichiro},
	doi = {10.7566/JPSJ.92.081006},
	journal = {Journal of the Physical Society of Japan},
	number = {8},
	pages = {081006},
	title = {{Generalized Dzyaloshinskii-Moriya Interaction and Chirality-Induced Phenomena in Chiral Crystals}},
	volume = {92},
	year = {2023}}

@article{Zhang2026,
	author = {Zhang, Shuai and Huang, Zhiheng and Du, Muchen and Ying, Tianping and Du, Luojun and Zhang, Tiantian},
	issue = {2},
	journal = {Phys. Rev. B},
	month = {Jan},
	numpages = {10},
	pages = {024302},
	publisher = {American Physical Society},
	title = {{Comprehensive study of phonon chirality under symmetry constraints}},
	volume = {113},
	year = {2026}}

@article{Juraschek2025,
	author = {Juraschek, Dominik M. and Geilhufe, R. Matthias and Zhu, Hanyu and Basini, Martina and Baum, Peter and Baydin, Andrey and Chaudhary, Swati and Fechner, Michael and Flebus, Benedetta and Grissonnanche, Gael and Kirilyuk, Andrei I. and Lemeshko, Mikhail and Maehrlein, Sebastian F. and Mignolet, Maxime and Murakami, Shuichi and Niu, Qian and Nowak, Ulrich and Romao, Carl P. and Rostami, Habib and Satoh, Takuya and Spaldin, Nicola A. and Ueda, Hiroki and Zhang, Lifa},
	journal = {Nature Physics},
	number = {10},
	pages = {1532--1540},
	title = {Chiral phonons},
	volume = {21},
	year = {2025}}

@article{Bloom2024,
	author = {Bloom, Brian P. and Paltiel, Yossi and Naaman, Ron and Waldeck, David H.},
	journal = {Chem. Rev.},
	number = {4},
	pages = {1950-1991},
	title = {Chiral Induced Spin Selectivity},
	volume = {124},
	year = {2024}}

@article{Bousquet2025,
	author = {Bousquet, Eric and Fava, Mauro and Romestan, Zachary and G\'omez-Ortiz, Fernando and McCabe, Emma E and Romero, Aldo H},
	journal = {J. Phys.: Condens. Matter},
	month = {mar},
	number = {16},
	pages = {163004},
	publisher = {IOP Publishing},
	title = {Structural chirality and related properties in periodic inorganic solids: review and perspectives},
	volume = {37},
	year = {2025}}

@article{hattori2024effect,
	author = {Hattori, Kohei and Watanabe, Hikaru and Iguchi, Junta and Nomoto, Takuya and Arita, Ryotaro},
	doi = {10.1103/PhysRevB.110.014425},
	issue = {1},
	journal = {Phys. Rev. B},
	month = {Jul},
	numpages = {9},
	pages = {014425},
	publisher = {American Physical Society},
	title = {Effect of collective spin excitations on electronic transport in topological spin textures},
	volume = {110},
	year = {2024}}

@article{hattori2025nonlinear,
	author = {Hattori, Kohei and Watanabe, Hikaru and Arita, Ryotaro},
	doi = {10.1103/PhysRevB.111.174416},
	issue = {17},
	journal = {Phys. Rev. B},
	month = {May},
	numpages = {31},
	pages = {174416},
	publisher = {American Physical Society},
	title = {Nonlinear Hall effect driven by spin-charge-coupled motive force},
	volume = {111},
	year = {2025}}

@article{kirikoshi2025light,
	author = {Kirikoshi, Akimitsu and Hayami, Satoru},
	journal = {arXiv:2509.14241},
	title = {{Light-induced nonlinear Edelstein effect under ferroaxial ordering}}}

@article{yoshimi2025,
	journal = {arXiv:2408.04450},
	author = {Kosuke Yoshimi and Yusuke Kato and Yuta Suzuki and Shuntaro Sumita and Takuro Sato and Hiroshi M. Yamamoto and Yoshihiko Togawa and Hiroaki Kusunose and Jun-ichiro Kishine},
	title = {{Chirality-dependent spin polarization in metals: linear and quadratic responses}}}

@article{cheong2021permutable,
	author = {Cheong, Sang-Wook and Lim, Seongjoon and Du, Kai and Huang, Fei-Ting},
	doi = {10.1038/s41535-021-00346-1},
	journal = {npj Quantum Mater.},
	number = {1},
	pages = {58},
	publisher = {Nature Publishing Group},
	title = {{Permutable SOS (symmetry operational similarity)}},
	volume = {6},
	year = {2021}}

@article{cheong2022linking,
	author = {Cheong, Sang-Wook and Huang, Fei-Ting and Kim, Minhyong},
	doi = {10.1088/1361-6633/ac97aa},
	journal = {Rep. Prog. Phys.},
	pages = {124501},
	publisher = {IOP Publishing},
	title = {{Linking emergent phenomena and broken symmetries through one-dimensional objects and their dot/cross products}},
	volume = {85},
	year = {2022}}

@article{kusunose2024emergence,
	author = {Kusunose, Hiroaki and Kishine, Jun-ichiro and Yamamoto, Hiroshi M},
	doi = {10.1063/5.0214919},
	issn = {0003-6951},
	journal = {Appl. Phys. Lett.},
	month = {06},
	number = {26},
	pages = {260501},
	publisher = {AIP Publishing},
	title = {{Emergence of chirality from electron spins, physical fields, and material-field composites}},
	volume = {124},
	year = {2024}}

@article{hayami2025chirality,
	author = {Hayami, Satoru and Oiwa, Rikuto and Inda, Akane},
	doi = {https://doi.org/10.7566/JPSJ.94.123702},
	journal = {J. Phys. Soc. Jpn.},
	number = {12},
	pages = {123702},
	publisher = {The Physical Society of Japan},
	title = {{Chirality/Axiality-Induced Axiality/Chirality via Surface Polarization}},
	volume = {94},
	year = {2025}}

@article{edelstein1990spin,
	author = {Edelstein, Victor M},
	doi = {10.1016/0038-1098(90)90963-C},
	journal = {Solid State Commun.},
	number = {3},
	pages = {233--235},
	publisher = {Elsevier},
	title = {{Spin polarization of conduction electrons induced by electric current in two-dimensional asymmetric electron systems}},
	volume = {73},
	year = {1990}}

@article{Miki_PhysRevLett.134.226401,
	author = {Miki, Tatsuya and Ikeda, Hiroaki and Suzuki, Michi-To and Hoshino, Shintaro},
	doi = {10.1103/PhysRevLett.134.226401},
	issue = {22},
	journal = {Phys. Rev. Lett.},
	month = {Jun},
	numpages = {7},
	pages = {226401},
	publisher = {American Physical Society},
	title = {{Quantification of Electronic Asymmetry: Chirality and Axiality in Solids}},
	volume = {134},
	year = {2025}}

@article{ishitobi2025purely,
  title = {Purely Electronic Chirality without Structural Chirality},
  author = {Ishitobi, Takayuki and Hattori, Kazumasa},
  journal = {Phys. Rev. Lett.},
  volume = {136},
  issue = {5},
  pages = {056402},
  numpages = {6},
  year = {2026},
  month = {Feb},
  publisher = {American Physical Society},
  doi = {10.1103/dc1q-xzbd},
  url = {https://link.aps.org/doi/10.1103/dc1q-xzbd}
}

@article{Hayami_PhysRevLett.122.147602,
	author = {Hayami, Satoru and Yanagi, Yuki and Kusunose, Hiroaki and Motome, Yukitoshi},
	doi = {10.1103/PhysRevLett.122.147602},
	issue = {14},
	journal = {Phys. Rev. Lett.},
	month = {Apr},
	numpages = {6},
	pages = {147602},
	publisher = {American Physical Society},
	title = {{Electric Toroidal Quadrupoles in the Spin-Orbit-Coupled Metal {${\mathrm{Cd}}_{2}{\mathrm{Re}}_{2}{\mathrm{O}}_{7}$}}},
	volume = {122},
	year = {2019}}

@book{Barron_mol-light-scattering,
	author = {L. D. Barron},
	chapter = {1.9.5},
	edition = {Second},
	pages = {39},
	publisher = {Cambridge University Press, Cambridge, U.K.},
	title = {{Molecular Light Scattering and Optical Activity}},
	year = {2004}}

@article{L.D.Barron_1986_true-chirality,
	author = {L. D. Barron},
	doi = {https://doi.org/10.1016/0009-2614(86)80035-5},
	issn = {0009-2614},
	journal = {Chem. Phys. Lett.},
	number = {5},
	pages = {423-427},
	title = {{True and false chirality and parity violation}},
	volume = {123},
	year = {1986}}

@article{B.Gohler_nat_2011_CISS,
	author = {B. G{\"o}hler and V. Hamelbeck and T. Z. Markus and M. Kettner and G. F. Hanne and Z. Vager and R. Naaman and H. Zacharias},
	doi = {10.1126/science.1199339},
	journal = {Science},
	number = {6019},
	pages = {894-897},
	title = {{Spin Selectivity in Electron Transmission Through Self-Assembled Monolayers of Double-Stranded $\mathrm{DNA}$}},
	volume = {331},
	year = {2011}}

@article{Xie_2011_CISS,
	author = {Xie, Zouti and Markus, Tal Z. and Cohen, Sidney R. and Vager, Zeev and Gutierrez, Rafael and Naaman, Ron},
	doi = {10.1021/nl2021637},
	journal = {Nano Letters},
	number = {11},
	pages = {4652-4655},
	title = {{Spin Specific Electron Conduction through DNA Oligomers}},
	volume = {11},
	year = {2011}}

@article{O.Ben_nat_2017_CISS,
	author = {Ben Dor, Oren and Yochelis, Shira and Radko, Anna and Vankayala, Kiran and Capua, Eyal and Capua, Amir and Yang, See-Hun and Baczewski, Lech Tomasz and Parkin, Stuart Stephen Papworth and Naaman, Ron and others},
	journal = {Nat. Commun.},
	number = {1},
	pages = {14567},
	publisher = {Nature Publishing Group UK London},
	title = {{Magnetization switching in ferromagnets by adsorbed chiral molecules without current or external magnetic field}},
	volume = {8},
	year = {2017}}

@article{K.Michaeli_PNAS_2019_CISS,
	author = {Karen Michaeli and David N. Beratan and David H. Waldeck and Ron Naaman},
	doi = {10.1073/pnas.1816956116},
	journal = {PNAS},
	number = {13},
	pages = {5931-5936},
	title = {{Voltage-induced long-range coherent electron transfer through organic molecules}},
	volume = {116},
	year = {2019}}

@article{Suda_natcom_2019_CISS,
	author = {Suda, Masayuki and Thathong, Yuranan and Promarak, Vinich and Kojima, Hirotaka and Nakamura, Masakazu and Shiraogawa, Takafumi and Ehara, Masahiro and Yamamoto, Hiroshi M.},
	date = {2019/06/05},
	doi = {10.1038/s41467-019-10423-6},
	id = {Suda2019},
	isbn = {2041-1723},
	journal = {Nat. Commun.},
	number = {1},
	pages = {2455},
	title = {{Light-driven molecular switch for reconfigurable spin filters}},
	volume = {10},
	year = {2019}}

@article{Haipeng_2019_CISS,
	author = {Haipeng Lu and Jingying Wang and Chuanxiao Xiao and Xin Pan and Xihan Chen and Roman Brunecky and Joseph J. Berry and Kai Zhu and Matthew C. Beard and Zeev Valy Vardeny},
	doi = {10.1126/sciadv.aay0571},
	journal = {Sci. Adv.},
	number = {12},
	pages = {eaay0571},
	title = {{Spin-dependent charge transport through 2D chiral hybrid lead-iodide perovskites}},
	volume = {5},
	year = {2019}}

@article{Naaman2020_2020_chiral_molecules_spin_selectivity,
	author = {Naaman, R. and Paltiel, Y. and Waldeck, D. H.},
	doi = {10.1021/acs.jpclett.0c00474},
	journal = {J. Phys. Chem. Lett.},
	number = {9},
	pages = {3660-3666},
	title = {Chiral Molecules and the Spin Selectivity Effect},
	volume = {11},
	year = {2020}}

@article{A.Inui2020prl_CrNb3S6,
	author = {Inui, Akito and Aoki, Ryuya and Nishiue, Yuki and Shiota, Kohei and Kousaka, Yusuke and Shishido, Hiroaki and Hirobe, Daichi and Suda, Masayuki and Ohe, Jun-ichiro and Kishine, Jun-ichiro and Yamamoto, Hiroshi M. and Togawa, Yoshihiko},
	doi = {10.1103/PhysRevLett.124.166602},
	issue = {16},
	journal = {Phys. Rev. Lett.},
	month = {Apr},
	numpages = {6},
	pages = {166602},
	publisher = {American Physical Society},
	title = {{Chirality-Induced Spin-Polarized State of a Chiral Crystal ${\mathrm{CrNb}}_{3}{\mathrm{S}}_{6}$}},
	volume = {124},
	year = {2020}}

@article{R.Neeman_ACR_2020_CISS,
	author = {Naaman, Ron and Paltiel, Yossi and Waldeck, David H.},
	doi = {10.1021/acs.accounts.0c00485},
	journal = {Acc. Chem. Res.},
	number = {11},
	pages = {2659-2667},
	title = {Chiral Induced Spin Selectivity Gives a New Twist on Spin-Control in Chemistry},
	volume = {53},
	year = {2020}}

@article{D.H.Waldeck2021aplmat_CISS,
	author = {Waldeck, D. H. and Naaman, R. and Paltiel, Y.},
	doi = {10.1063/5.0049150},
	issn = {2166-532X},
	journal = {APL Mater.},
	month = {04},
	number = {4},
	pages = {040902},
	title = {{The spin selectivity effect in chiral materials}},
	volume = {9},
	year = {2021}}

@article{Y.Nabei_APL_2020_CrNb3S6,
	author = {Nabei, Yoji and Hirobe, Daichi and Shimamoto, Yusuke and Shiota, Kohei and Inui, Akito and Kousaka, Yusuke and Togawa, Yoshihiko and Yamamoto, Hiroshi M.},
	doi = {10.1063/5.0017882},
	issn = {0003-6951},
	journal = {Appl. Phys. Lett.},
	month = {08},
	number = {5},
	pages = {052408},
	title = {{Current-induced bulk magnetization of a chiral crystal CrNb3S6}},
	volume = {117},
	year = {2020}}

@article{K.Shiota2021prl_disilicide_CISS,
	author = {Shiota, Kohei and Inui, Akito and Hosaka, Yuta and Amano, Ryoga and \ifmmode \bar{O}\else \={O}\fi{}nuki, Yoshichika and Hedo, Masato and Nakama, Takao and Hirobe, Daichi and Ohe, Jun-ichiro and Kishine, Jun-ichiro and Yamamoto, Hiroshi M. and Shishido, Hiroaki and Togawa, Yoshihiko},
	doi = {10.1103/PhysRevLett.127.126602},
	issue = {12},
	journal = {Phys. Rev. Lett.},
	month = {Sep},
	numpages = {5},
	pages = {126602},
	publisher = {American Physical Society},
	title = {{Chirality-Induced Spin Polarization over Macroscopic Distances in Chiral Disilicide Crystals}},
	volume = {127},
	year = {2021}}

@article{H.Shishido_apl_2021_NbSi2-TaSi2,
	author = {Shishido, Hiroaki and Sakai, Rei and Hosaka, Yuta and Togawa, Yoshihiko},
	doi = {10.1063/5.0074293},
	issn = {0003-6951},
	journal = {Appl. Phys. Lett.},
	month = {11},
	number = {18},
	pages = {182403},
	title = {{Detection of chirality-induced spin polarization over millimeters in polycrystalline bulk samples of chiral disilicides NbSi2 and TaSi2}},
	volume = {119},
	year = {2021}}

@article{F.Evers_admat_2022_CISS,
	author = {Evers, Ferdinand and Aharony, Amnon and Bar-Gill, Nir and Entin-Wohlman, Ora and Hedeg\r{a}rd, Per and Hod, Oded and Jelinek, Pavel and Kamieniarz, Grzegorz and Lemeshko, Mikhail and Michaeli, Karen and Mujica, Vladimiro and Naaman, Ron and Paltiel, Yossi and Refaely-Abramson, Sivan and Tal, Oren and Thijssen, Jos and Thoss, Michael and van Ruitenbeek, Jan M. and Venkataraman, Latha and Waldeck, David H. and Yan, Binghai and Kronik, Leeor},
	doi = {https://doi.org/10.1002/adma.202106629},
	journal = {Adv. Mater.},
	number = {13},
	pages = {2106629},
	title = {{Theory of Chirality Induced Spin Selectivity: Progress and Challenges}},
	volume = {34},
	year = {2022}}

@article{A.Kato2022prb_CISS,
	author = {Kato, Akihito and Yamamoto, Hiroshi M. and Kishine, Jun-ichiro},
	doi = {10.1103/PhysRevB.105.195117},
	issue = {19},
	journal = {Phys. Rev. B},
	month = {May},
	numpages = {9},
	pages = {195117},
	publisher = {American Physical Society},
	title = {{Chirality-induced spin filtering in pseudo Jahn-Teller molecules}},
	volume = {105},
	year = {2022}}

@article{Nakajima23,
	author = {Nakajima, R. and Hirobe, D. and Kawaguchi, G. and Nabei, Y. and Sato, T. and Narushima, T. and Okamoto, H. and Yamamoto, H. M.},
	date = {2023/01/01},
	doi = {10.1038/s41586-022-05589-x},
	id = {Nakajima2023},
	isbn = {1476-4687},
	journal = {Nature},
	number = {7944},
	pages = {479--484},
	title = {{Giant spin polarization and a pair of antiparallel spins in a chiral superconductor}},
	volume = {613},
	year = {2023}}

@article{K.Ohe_2024_quartz_CISS,
	author = {Ohe, Kazuki and Shishido, Hiroaki and Kato, Masaki and Utsumi, Shoyo and Matsuura, Hiroyasu and Togawa, Yoshihiko},
	doi = {10.1103/PhysRevLett.132.056302},
	issue = {5},
	journal = {Phys. Rev. Lett.},
	month = {Jan},
	numpages = {7},
	pages = {056302},
	publisher = {American Physical Society},
	title = {{Chirality-Induced Selectivity of Phonon Angular Momenta in Chiral Quartz Crystals}},
	volume = {132},
	year = {2024}}

@article{Bloom_chemrev_2024_CISS,
	author = {Bloom, Brian P. and Paltiel, Yossi and Naaman, Ron and Waldeck, David H.},
	doi = {10.1021/acs.chemrev.3c00661},
	journal = {Chem. Rev.},
	number = {4},
	pages = {1950-1991},
	title = {{Chiral Induced Spin Selectivity}},
	volume = {124},
	year = {2024}}

@article{K.Banerjee-Ghosh_science_2018_Separation_of_enantiomers,
	author = {Koyel Banerjee-Ghosh and Oren Ben Dor and Francesco Tassinari and Eyal Capua and Shira Yochelis and Amir Capua and See-Hun Yang and Stuart S. P. Parkin and Soumyajit Sarkar and Leeor Kronik and Lech Tomasz Baczewski and Ron Naaman and Yossi Paltiel},
	doi = {10.1126/science.aar4265},
	journal = {Science},
	number = {6395},
	pages = {1331-1334},
	title = {{Separation of enantiomers by their enantiospecific interaction with achiral magnetic substrates}},
	volume = {360},
	year = {2018}}

@article{Safari_admat_2024_enantioselective_adsorption,
	author = {Safari, Mohammad Reza and Matthes, Frank and Caciuc, Vasile and Atodiresei, Nicolae and Schneider, Claus M. and Ernst, Karl-Heinz and B\"{u}rgler, Daniel E.},
	doi = {https://doi.org/10.1002/adma.202308666},
	journal = {Adv. Mater.},
	number = {14},
	pages = {2308666},
	title = {{Enantioselective Adsorption on Magnetic Surfaces}},
	volume = {36},
	year = {2024}}

@article{S.Miwa_sciadv_2025_spin-polarization_enantioselectivity,
	author = {Shinji Miwa and Tatsuya Yamamoto and Takashi Nagata and Shoya Sakamoto and Kenta Kimura and Masanobu Shiga and Weiguang Gao and Hiroshi M. Yamamoto and Keiichi Inoue and Taishi Takenobu and Takayuki Nozaki and Tatsuhiko Ohto},
	doi = {10.1126/sciadv.adv5220},
	journal = {Sci. Adv.},
	number = {44},
	pages = {eadv5220},
	title = {Spin polarization driven by molecular vibrations leads to enantioselectivity in chiral molecules},
	volume = {11},
	year = {2025}}

@article{hoshino2023spin-derived,
	author = {Hoshino, Shintaro and Suzuki, Michi-To and Ikeda, Hiroaki},
	doi = {10.1103/PhysRevLett.130.256801},
	issue = {25},
	journal = {Phys. Rev. Lett.},
	month = {Jun},
	numpages = {6},
	pages = {256801},
	publisher = {American Physical Society},
	title = {{Spin-Derived Electric Polarization and Chirality Density Inherent in Localized Electron Orbitals}},
	volume = {130},
	year = {2023}}

@article{Oiwa_PhysRevLett.129.116401,
	author = {Oiwa, Rikuto and Kusunose, Hiroaki},
	doi = {10.1103/PhysRevLett.129.116401},
	issue = {11},
	journal = {Phys. Rev. Lett.},
	month = {Sep},
	numpages = {6},
	pages = {116401},
	publisher = {American Physical Society},
	title = {{Rotation, Electric-Field Responses, and Absolute Enantioselection in Chiral Crystals}},
	volume = {129},
	year = {2022}}

@article{J.Kishine_IJC_2022_G0,
	author = {Kishine, Jun-ichiro and Kusunose, Hiroaki and Yamamoto, Hiroshi M.},
	doi = {https://doi.org/10.1002/ijch.202200049},
	journal = {Isr. J. Chem.},
	number = {11-12},
	pages = {e202200049},
	title = {{On the Definition of Chirality and Enantioselective Fields}},
	volume = {62},
	year = {2022}}

@article{hayami2023chiral,
	author = {Hayami , Satoru and Kusunose, Hiroaki},
	doi = {https://doi.org/10.7566/JPSJ.92.113704},
	journal = {J. Phys. Soc. Jpn.},
	pages = {113704},
	title = {{Chiral charge as hidden order parameter in URu2Si2}},
	volume = {92},
	year = {2023}}

@article{hayami2024analysis,
	author = {Hayami, Satoru and Yambe, Ryota and Kusunose, Hiroaki},
	doi = {https://doi.org/10.7566/JPSJ.93.043702},
	journal = {J. Phys. Soc. Jpn.},
	number = {4},
	pages = {043702},
	publisher = {The Physical Society of Japan},
	title = {Analysis of photo-induced chirality and magnetic toroidal moment based on Floquet formalism},
	volume = {93},
	year = {2024}}

@article{inda2024quantification,
	author = {Inda, A. and Oiwa, R. and Hayami, S. and Yamamoto, H. M. and Kusunose, H.},
	doi = {10.1063/5.0204254},
	issn = {0021-9606},
	journal = {J. Chem. Phys.},
	month = {05},
	number = {18},
	pages = {184117},
	title = {{Quantification of chirality based on electric toroidal monopole}},
	volume = {160},
	year = {2024}}

@article{oiwa2025prr_te_se,
	author = {Oiwa, Rikuto and Kusunose, Hiroaki},
	doi = {10.1103/1zq8-pqh8},
	issue = {3},
	journal = {Phys. Rev. Res.},
	month = {Sep},
	numpages = {15},
	pages = {033250},
	publisher = {American Physical Society},
	title = {{Predominant electronic order parameter for structural chirality: Role of spinless electronic toroidal multipoles in Te and Se}},
	volume = {7},
	year = {2025}}

@article{Furukawa2017,
	author = {Furukawa, Tetsuya and Shimokawa, Yuri and Kobayashi, Kaya and Itou, Tetsuaki},
	date = {2017/10/16},
	doi = {10.1038/s41467-017-01093-3},
	id = {Furukawa2017},
	isbn = {2041-1723},
	journal = {Nat. Commun.},
	number = {1},
	pages = {954},
	title = {{Observation of current-induced bulk magnetization in elemental tellurium}},
	volume = {8},
	year = {2017}}

@article{T.Yoda_sr_2015_Edelstein,
	author = {Yoda, Taiki and Yokoyama, Takehito and Murakami, Shuichi},
	date = {2015/07/09},
	doi = {10.1038/srep12024},
	id = {Yoda2015},
	isbn = {2045-2322},
	journal = {Sci. Rep.},
	number = {1},
	pages = {12024},
	title = {{Current-induced Orbital and Spin Magnetizations in Crystals with Helical Structure}},
	volume = {5},
	year = {2015}}

@article{T.Furukawa_prr_2021_Edelstein,
	author = {Furukawa, Tetsuya and Watanabe, Yuta and Ogasawara, Naoki and Kobayashi, Kaya and Itou, Tetsuaki},
	doi = {10.1103/PhysRevResearch.3.023111},
	issue = {2},
	journal = {Phys. Rev. Res.},
	month = {May},
	numpages = {14},
	pages = {023111},
	publisher = {American Physical Society},
	title = {{Current-induced magnetization caused by crystal chirality in nonmagnetic elemental tellurium}},
	volume = {3},
	year = {2021}}

@article{hayami2020prb_bottom-up,
	author = {Hayami, Satoru and Yanagi, Yuki and Kusunose, Hiroaki},
	doi = {10.1103/PhysRevB.102.144441},
	issue = {14},
	journal = {Phys. Rev. B},
	month = {Oct},
	numpages = {24},
	pages = {144441},
	publisher = {American Physical Society},
	title = {{Bottom-up design of spin-split and reshaped electronic band structures in antiferromagnets without spin-orbit coupling: Procedure on the basis of augmented multipoles}},
	volume = {102},
	year = {2020}}

@article{R.Oiwa2022jpsj_nonl-respo,
	author = {Oiwa ,Rikuto and Kusunose ,Hiroaki},
	doi = {10.7566/JPSJ.91.014701},
	journal = {J. Phys. Soc. Jpn},
	number = {1},
	pages = {014701},
	title = {{Systematic Analysis Method for Nonlinear Response Tensors}},
	volume = {91},
	year = {2022}}

@article{Yue:22,
	author = {Lun Yue and Mette B. Gaarde},
	doi = {10.1364/JOSAB.448602},
	journal = {J. Opt. Soc. Am. B},
	month = {Feb},
	number = {2},
	pages = {535--555},
	publisher = {Optica Publishing Group},
	title = {{Introduction to theory of high-harmonic generation in solids: tutorial}},
	volume = {39},
	year = {2022}}

@article{Kusunose_PhysRevB.107.195118,
	author = {Kusunose, Hiroaki and Oiwa, Rikuto and Hayami, Satoru},
	doi = {10.1103/PhysRevB.107.195118},
	issue = {19},
	journal = {Phys. Rev. B},
	month = {May},
	numpages = {14},
	pages = {195118},
	publisher = {American Physical Society},
	title = {{Symmetry-adapted modeling for molecules and crystals}},
	volume = {107},
	year = {2023}}

@article{hayami2024unified,
	author = {Hayami, Satoru and Kusunose, Hiroaki},
	doi = {https://doi.org/10.7566/JPSJ.93.072001},
	journal = {J. Phys. Soc. Jpn.},
	number = {7},
	pages = {072001},
	publisher = {The Physical Society of Japan},
	title = {{Unified description of electronic orderings and cross correlations by complete multipole representation}},
	volume = {93},
	year = {2024}}

@article{kusunose2020jpsj_completebasis,
	author = {Kusunose, Hiroaki and Oiwa, Rikuto and Hayami, Satoru},
	doi = {10.7566/JPSJ.89.104704},
	journal = {J. Phys. Soc. Jpn.},
	number = {10},
	pages = {104704},
	publisher = {The Physical Society of Japan},
	title = {{Complete Multipole Basis Set for Single-Centered Electron Systems}},
	volume = {89},
	year = {2020}}

@article{hayami2018prb_Classification_of_atomic-scale_multipoles,
	author = {Hayami, Satoru and Yatsushiro, Megumi and Yanagi, Yuki and Kusunose, Hiroaki},
	doi = {10.1103/PhysRevB.98.165110},
	issue = {16},
	journal = {Phys. Rev. B},
	month = {Oct},
	numpages = {35},
	pages = {165110},
	publisher = {American Physical Society},
	title = {{Classification of atomic-scale multipoles under crystallographic point groups and application to linear response tensors}},
	volume = {98},
	year = {2018}}

@article{yatsushiro2021prb_122,
	author = {Yatsushiro, Megumi and Kusunose, Hiroaki and Hayami, Satoru},
	doi = {10.1103/PhysRevB.104.054412},
	issue = {5},
	journal = {Phys. Rev. B},
	month = {Aug},
	numpages = {41},
	pages = {054412},
	publisher = {Am. Phys. Soc.},
	title = {{Multipole classification in 122 magnetic point groups for unified understanding of multiferroic responses and transport phenomena}},
	volume = {104},
	year = {2021}}
\end{document}